\documentclass[aps,prl,notitlepage,amsmath,amssymb,amsfonts,longbibliography,superscriptaddress,twocolumn]{revtex4-2}

\makeatletter
\def\@hangfrom@section#1#2#3{\normalsize\@hangfrom{#1#2}#3}
\def\@hangfroms@section#1#2{\normalsize#1#2}
\makeatother

\usepackage{graphicx,soul}
\usepackage{dcolumn}
\usepackage{bm}
\usepackage{bbm} 

\usepackage{dsfont}
\usepackage{enumerate}

\usepackage[dvipsnames, x11names]{xcolor}
\usepackage[colorlinks=true]{hyperref}
\hypersetup{
    colorlinks=true,
    citecolor=blue,
    linkcolor=red,
    urlcolor=RoyalBlue,
    anchorcolor=Purple
}

\usepackage[percent]{overpic}
\usepackage{tabularx}
\usepackage{multirow}
\usepackage{CJK}
\usepackage{orcidlink}

\usepackage{physics}

\newcommand{\vk}{\mathbf k}
\newcommand{\vR}{\mathbf R}
\newcommand{\udis}{\mathbf u}
\newcommand{\Id}{\mathbb I}
\newcommand{\utv}{\widetilde{\mathbf u}}
\newcommand{\ptv}{\widetilde{\mathbf p}}

\newcommand{\bluefour}[1]{\textcolor{Blue4}{#1}}

\def\be{\begin{equation}}
\def\ee{\end{equation}}
\def\bea{\begin{eqnarray}}
\def\eea{\end{eqnarray}}

\begin{document}

\title{\texorpdfstring{\resizebox{\textwidth}{!}{Intertwined complex orbital order and ferrimagnetism in honeycomb molybdate Fe$_2$Mo$_3$O$_8$}}{Fe2Mo3O8}}
\author{Hong-Hao Song}
\affiliation{International Center for Quantum Materials, School of Physics, 
Peking University, Beijing 100871, China}
\author{Gang v.~Chen}
\email{chenxray@pku.edu.cn}
\affiliation{International Center for Quantum Materials, School of Physics, 
Peking University, Beijing 100871, China}
\affiliation{Collaborative Innovation Center of Quantum Matter, 100871, Beijing, China}
\affiliation{Beijing Key Laboratory of Quantum Devices, Peking University, Beijing 100871, China}
\date{\today}


\begin{abstract}
Motivated by the recent experimental interest in honeycomb molybdates A$_2$Mo$_3$O$_8$,  
we return and develop the minimal models to capture the essential physics for Fe$_2$Mo$_3$O$_8$. 
It is pointed out that, the magnetism in these materials arises from the mutual induction
between the spin-orbit coupling of the tetrahedral sublattice with active orbitals
and the superexchange with the spins from the octahedral sublattice. 
Essentially, the superexchange between the tetrahedral and octahedral sublattices directly
drives the ferrimagnetism, and the residual spin-orbit coupling on the tetrahedral 
sublattice induces the complex orbital order. 
The hybrid magnon–phonon excitations are further explored together 
with their resulting Berry curvatures and thermal Hall effect.
We discuss the experimental relevance with
Fe$_2$Mo$_3$O$_8$, and the framework can be extended to FeNiMo$_3$O$_8$, 
Ni$_2$Mo$_3$O$_8$ and other isostructural magnets. 
\end{abstract}

\maketitle

\noindent\bluefour{\it Introduction.}---Understanding the physical degrees of freedom 
encoded in the microscopic 
energy-level structure of a quantum many-body system 
is among the most fundamental questions in condensed matter physics, 
for it constitutes the essential starting point from which the effective models 
are constructed and emergent phenomena are explored. 
The spin, orbital, and valley degrees of freedom of electrons in solids, 
for instance, underlie a rich landscape of physics ranging from magnetism 
to correlated topological phases~\cite{kugel1982,schaibley2016,rau2016}, 
and only after their energy scales and symmetry properties are carefully 
identified can reliable low-energy theories be established. The same line 
of process governs the study of ultracold atomic gases, 
where the internal (hyperfine) energy levels of atoms serve as 
synthetic degrees of freedom~\cite{celi2014}. A deep understanding of these atomic levels, 
their coupling strengths, scattering properties, and symmetry structure, 
is a prerequisite for the exquisite control achieved in modern atomic physics, 
as exemplified by magnetically tunable Feshbach resonances~\cite{chin2010} 
and the engineering of artificial gauge fields~\cite{dalibard2011}.

Along this line of practice, spin-orbit-coupled Mott insulators offer a distinct platform 
in which genuinely new physical degrees of freedom are introduced~\cite{witczak2014correlated}. 
In these materials, 
the strong spin-orbit coupling (SOC) entangles the spin and orbital degrees of freedom of a local moment, 
and the ensuing Mott insulating state stabilizes an effective total angular momentum 
(e.g., a ${j_\mathrm{eff} = 1/2}$ doublet) that 
behaves as a new pseudospin~\cite{kim2008,Chen2008,jackeli2009}. 
This emergent degree of freedom, simultaneously spin and orbital in character, 
cannot be reduced to either ingredient alone, and it underpins a wealth of phenomena 
ranging from Kitaev quantum spin liquids~\cite{kitaev2006,jackeli2009} 
to unconventional multipolar orders~\cite{chen2010,Chen2011}. 
So far, however, such physics has been explored predominantly 
in $4d/5d$ transition-metal and $4f$ rare-earth systems, where 
the sizable SOC naturally justifies the importance of the atomic multiplet~\cite{witczak2014correlated}. 
The multiplet structure of $3d$ transition-metal ions, 
by contrast, has received comparatively little attention,  
largely because their SOC is conventionally regarded as weak. 
Yet many $3d$ compounds in fact possess a remarkably rich multiplet structure 
that can give rise to unexpected new physics~\cite{maekawa2004physics,chen2009,Chen2010prb}. 
Moreover, the SOC of $3d$ ions 
can play a crucial role in specific settings, as exemplified 
in the context of Co-based magnets~\cite{liu2018,Sano2018}, 
and as realized in many other $3d$ systems where sizable spin-orbit effects have been identified~\cite{PhysRevB.100.045103,PhysRevResearch.5.L032042}. 
Whereas recent efforts have sought non-Heisenberg spin interactions indirectly, 
for instance through the SOC of ligand ions in materials 
such as CrI$_3$~\cite{lado2017} and NiI$_2$~\cite{amoroso2020}, 
it is arguably more natural and direct to investigate these phenomena  
in $3d$ transition-metal compounds where the SOC of the magnetic 
ion itself is substantial.

In this Letter, we take the compound Fe$_2\text{Mo}_3\text{O}_8$ as a concrete and experimentally accessible starting point~\cite{ideue2017}. Like other compounds in this family, $\text{Fe}_2\text{Mo}_3\text{O}_8$ 
crystallizes in a polar hexagonal structure where the magnetic Fe ions form a honeycomb lattice~\cite{wang2015,bao2023}. 
A distinctive feature is that the two honeycomb sublattices 
are crystallographically inequivalent, hosting the Fe$^{2+}$ ions in two different coordinations, 
one sublattice is tetrahedrally and the other octahedrally coordinated~\cite{stanislavchuk2020}. 
This geometry simultaneously realizes several favourable ingredients, 
a low-dimensional frustrated magnetic network, 
a built-in electric polarization that couples lattice and magnetism~\cite{wang2015,kurumaji2015}, and most relevant here, 
a rich local multiplet structure tailored by the inequivalent local environments~\cite{stanislavchuk2020}. 
More specifically, the {Fe}$^{2+}$ ion in the tetrahedral environment 
supports a lower $e_1 = \{\vert d_{x^2-y^2} \rangle, \vert d_{xy} \rangle\}$ orbital degeneracy with
$z$ axis parallel to the $c$ axis besides having the total spin $S=2$. 
Remarkably, SOC is active even at
the first order. 
The octahedral Fe ion, however, has a fully quenched orbital and thus has a pure total spin $S=2$. 
The inter-sublattice superexchange directly drives the anticollinear ferrimagnetic order, 
which simultaneously acts as an internal magnetic field that induces the complex orbital
order on the tetrahedral sublattice. 
The excitations of the intertwined orbital order and ferrimagnetism are further corroborated 
with the spectrum and thermal Hall transports.  
This family of materials therefore provides an ideal setting to examine 
how the internal energy-level structure of $3d$ moments translates  
into emergent low-energy degrees of freedom and magnetic phenomena.

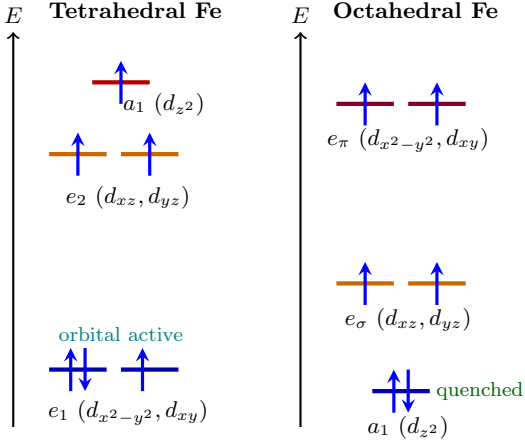
\begin{figure}[t]
\centering
\begin{tikzpicture}[scale=0.95, every node/.style={transform shape}]

    \begin{scope}[shift={(0,0)}]
        \node[align=center, font=\bfseries] at (1.7, 5.8) {Tetrahedral $\text{Fe}$};
        
        \draw[->, thick] (0, 0) -- (0, 5.5) node[above] {$E$};
        
        \draw[ultra thick, color=red!70!black] (1.1, 4.8) -- (1.9, 4.8) node[right, black] {};
        \node[font=\footnotesize, text=red!70!black] at (2, 5.1) {};
                        \node[ black] at (2.1, 4.5) {$a_1\ (d_{z^2})$};

        \draw[ultra thick, color=orange!80!black] (0.5, 3.8) -- (1.3, 3.8);
        \draw[ultra thick, color=orange!80!black] (1.5, 3.8) -- (2.3, 3.8) node[right, black] {};
                \node[ black] at (1.6, 3.2) {$e_2\ (d_{xz}, d_{yz})$};

        \draw[ultra thick, color=blue!70!black] (0.5, 0.8) -- (1.3, 0.8);
        \draw[ultra thick, color=blue!70!black] (1.5, 0.8) -- (2.3, 0.8) node[right, black] {};
        \node[ black] at (1.6, 0.2) {$e_1\ (d_{x^2-y^2}, d_{xy})$};
        
        \draw[->,>=stealth,line width=1.2pt,blue] (0.8, 0.5) -- (0.8, 1.1);
        \draw[<-,>=stealth,line width=1.2pt,blue] (1.0, 0.5) -- (1.0, 1.1);
        \draw[->,>=stealth,line width=1.2pt,blue] (1.8, 0.5) -- (1.8, 1.1);
        \node[teal, font=\footnotesize] at (1.5, 1.3) {orbital active};
        
        \draw[->,>=stealth,line width=1.2pt,blue] (1.5, 4.5) -- (1.5, 5.1);
                \draw[->,>=stealth,line width=1.2pt,blue] (0.9, 3.5) -- (0.9, 4.1);
                                \draw[->,>=stealth,line width=1.2pt,blue] (1.9, 3.5) -- (1.9, 4.1);

     \end{scope}

    \begin{scope}[shift={(4,0)}]
        \node[align=center, font=\bfseries] at (1.6, 5.8) {Octahedral Fe};
        
        \draw[->, thick] (0, 0) -- (0, 5.5) node[above] {$E$};
        
        \draw[ultra thick, color=purple!70!black] (0.5, 4.5) -- (1.3, 4.5);
        \draw[ultra thick, color=purple!70!black] (1.5, 4.5) -- (2.3, 4.5) node[right, black] {};
                 \node[ black] at (1.5, 1.5){$e_\sigma\ (d_{xz}, d_{yz})$};

        \draw[ultra thick, color=orange!80!black] (0.5, 2.0) -- (1.3, 2.0);
        \draw[ultra thick, color=orange!80!black] (1.5, 2.0) -- (2.3, 2.0) node[right, black] {};
                 \node[ black] at (1.5, 4.0){$e_\pi\ (d_{x^2-y^2}, d_{xy})$};

        \draw[ultra thick, color=blue!70!black] (1.0, 0.5) -- (1.8, 0.5) node[right, black] {};
         \node[ black] at (1.5, 0){$a_1\ (d_{z^2})$};

        \draw[->,>=stealth,line width=1.2pt,blue] (1.3, 0.2) -- (1.3, 0.8);
        \draw[<-,>=stealth,line width=1.2pt,blue] (1.5, 0.2) -- (1.5, 0.8);
        \node[black!60!green, font=\footnotesize] at (2.5, 0.5) {quenched};
        
        \draw[->,>=stealth,line width=1.2pt,blue] (0.9, 1.7) -- (0.9, 2.3);
        \draw[->,>=stealth,line width=1.2pt,blue] (1.9, 1.7) -- (1.9, 2.3);
                \draw[->,>=stealth,line width=1.2pt,blue] (0.9, 4.2) -- (0.9, 4.8);
        \draw[->,>=stealth,line width=1.2pt,blue] (1.9, 4.2) -- (1.9, 4.8);

    \end{scope}
\end{tikzpicture}
\caption{Schematic representation of the single-electron $3d$ crystal field splitting for $\text{Fe}^{2+}$ ($3d^6$) ions at the tetrahedral and octahedral sites in $\text{Fe}_2\text{Mo}_3\text{O}_8$ under a unified $z \parallel c$-axis frame. At the tetrahedral site, one hole appears in the doubly-degenerate $e_1$ state, leading to an orbital pseudospin $\tau = 1/2$. 
Conversely, at the trigonal-elongated octahedral site, the local orbitals are quenched.}
\label{fig1}
\end{figure}

\noindent\bluefour{\it Local multiplets.}---On the tetrahedral sublattice
of Fe$_2$Mo$_3$O$_8$, the crystal field splits the $d$ manifold into a lower $e_1$ doublet, 
a middle $e_2$ doublet, and an upper $a_1$ singlet (see Fig.~\ref{fig1})~\cite{reschke2020}. 
Under the frame where the $z$-axis aligns with the crystallographic $c$-axis, 
the doubly-degenerate $e_1$ manifold is explicitly composed of the 
$\{|d_{x^2-y^2}\rangle, |d_{xy}\rangle\}$ orbitals. 
The $3d^6$ configuration leaves one hole in this $e_1$ doublet, 
which hosts an active orbital degree of freedom represented by 
an orbital pseudospin $\tau = 1/2$ with $\tau^z=1/2$ ($-1/2$) for the hole
occupying the $x^2-y^2$ ($xy$) orbital. 
Within this real $e_1$ basis, 
the matrix elements of the axial orbital angular momentum operator ${L}_z$ 
are purely off-diagonal (${L}^z = 4\tau^y$), 
causing the first-order axial SOC to be active, 
with $H_{t} =- \lambda \sum_{ i \in t} S^z_i \tau^y_i $, 
where `${t}$' denotes the tetrahedral sublattice. Details are provided in the Supplemental Material.

On the octahedral Fe site, 
the local trigonal crystal field splits the $3d^6$ configuration differently (see Fig.~\ref{fig1}). 
Owing to the structural trigonal elongation along the $c$-axis, the apical oxygen ligands are pulled outward, drastically reducing the electrostatic repulsion along the $z$-direction. 
This projects the non-degenerate $a_1 = \{|d_{z^2}\rangle\}$ singlet to the bottom of the $t_{2g}$ manifold, leaving the two $E$-type doublets, $e^\pi$ and $e^\sigma$, at much higher energies. 
The electron occupation quenches the ground-state orbital degree of freedom~\cite{reschke2020}. The local moment behaves as a pure, orbitally-inert spin $S=2$ moment, where the SOC acts only as a second-order perturbation to yield a weak easy-axis single-ion anisotropy,
${H_{o} = -\sum_{i \in \text{o}} D_o \big(S_z^i\big)^2}$
with ${D_o > 0}$. Here `$o$' refers to the octahedral sublattice.

\noindent\bluefour{\it Superexchange.}---Having established the multiplet structure 
of the two Fe ions, we turn to the exchange coupling between them. 
The nearest-neighbour exchange interactions
 are mediated by the bridging oxygen through an Fe-O-Fe superexchange path (see Fig.~\ref{fig1})~\cite{anderson1959}. 
To capture them, one could construct a minimal hopping model for this Fe-O-Fe unit,  
retaining the full orbital content of the tetrahedral and octahedral sites. 
Here, there is only one oxygen site on this exchange path, one could 
project out the oxygen degrees of freedom in the second-order perturbation 
then yields an effective hopping Hamiltonian between the neighbouring Fe sites,
and then perform a complicated perturbative calculation to obtain the Kugel-Khomskii 
model~\cite{kugel1982}. Here, we avoid this boring and cumbersome calculation and instead rely on
a physical and microscopic reasoning.

To understand the exchange interaction between the two Fe sublattices, 
it is convenient to first ignore the SOC and decompose the moments 
into their orbitally resolved 
components. 
On the tetrahedral Fe site, the total ${S=2}$ spin moment is composed of two parts,
the half-filled $t_{2g}^3$ manifold (or equivalently $a_1^1e_2^2$) contributes to a spin ${S_{t}=3/2}$ that is orbitally inert, 
whereas the single unpaired electron in the doubly degenerate $e_1$ orbital contributes ${S_{e_1}=1/2}$ 
and carries the active orbital, which we have represented 
by an orbital pseudospin $\tau=1/2$. 
These two components are aligned by the intra-atomic Hund's coupling into the total spin
${\bf S}={\bf S}_{e_1} + {\bf S}_t$ with $S=2$.  
On the octahedral Fe site, 
the orbitals are fully quenched, leaving an orbitally inert ${S=2}$ moment. 
The superexchange between the two neighboring sites then separates naturally into two pieces. 
The exchange involving the three inert $t_{2g}$ electrons of 
the tetrahedral Fe site and the spin-only moment of the octahedral Fe site 
has to be a pure Heisenberg interaction. 
The remaining exchange involving the single $e_1$ electron of
the tetrahedral Fe also contains a Heisenberg component,  
but in addition acquires the typical Kugel–Khomskii terms 
in which the Heisenberg component is multiplied by the orbital pseudospin $\tau$ 
from the tetrahedral Fe site. 
Since these orbitally coupled terms involve only one of the four unpaired electrons 
and their hopping channels are orbitally selective, 
they are expected to be parametrically much weaker than the dominant Heisenberg part. 
Here there is no orbital-orbital exchange because the octahedral Fe site is not orbitally active. 
To the leading order, therefore, the nearest-neighbour superexchange 
interaction can be approximated by a Heisenberg interaction between the two $S=2$ 
spin moments, i.e.
 \begin{eqnarray}
H_{ex} = \sum_{i\in t}\sum_{j \in o} J\, {\bf S}_i \cdot {\bf S}_j + \cdots 
\end{eqnarray}
 with weak orbitally active corrections that set in only at a subleading scale. 
 The further neighbour exchange has been neglected. 
 As the system lacks the inversion symmetry, 
  ``$\cdots$'' here also includes the weak antisymmetric Dzyaloshinskii-Moriya (DM) interaction 
 that appears from one higher order perturbation than the superexchange~\cite{moriya1960}. 
 The full minimal model of the system is then written as 
\begin{eqnarray}
H_{min} = H_t + H_o + H_{ex} . 
\label{minimal}
\end{eqnarray}


\noindent\bluefour{\it {Ferrimagnetism and complex orbital order.}}---Remarkably, the minimal model has 
little frustration, and the ground state can be readily read off from the model. 
The two sublattices simply develop the anticollinearly aligned spin configuration. 
The easy-axis anisotropy of the octahedral sites prefers the spins to be aligned in the $\pm \hat{z}$
direction. 
For the tetrahedral sites, the onsite Ising-like SOC again favours the spins to be aligned in the $\pm \hat{z}$
direction in order to gain energy. 
In the end, as the two Fe moments are intrinsically different, despite that they are the anticollinearly aligned,
the antiferromagnetic state is a ferrimagnet. At the classical mean-field level, the spin and orbital configurations are then given as
\begin{equation}
\begin{aligned}
i \in t,\quad &\langle {\bf S}_i \rangle = 2 \hat{z},\quad
\langle \boldsymbol{\tau}_i \rangle = \frac{1}{2}\hat{y},\\
j \in o,\quad &\langle {\bf S}_j \rangle = -2\hat{z}.
\end{aligned}
\end{equation}
Here, the orbital pseudospin $\tau$ is directly polarized by $S^z$ at the tetrahedral sites via
the onsite SOC. The spin-orbital state on the tetrahedral site is 
given as
\begin{eqnarray}
|{S^z_i =2, \tau^y_i = \frac{1}{2} } \rangle \equiv 
|{S^z_i =2 }\rangle   \otimes \frac{1}{\sqrt{2}}|  d_{x^2-y^2} + \mathrm{i} d_{xy} \rangle_i,
\end{eqnarray}
where the orbital wavefunction is a complex one with the orbital angular momentum ${L^z_i = 2}$. 
All the tetrahedral sites have this complex orbital order.

\suppressfloats[t]
\begin{figure}[t]
\centering
\includegraphics[width=0.8\linewidth]{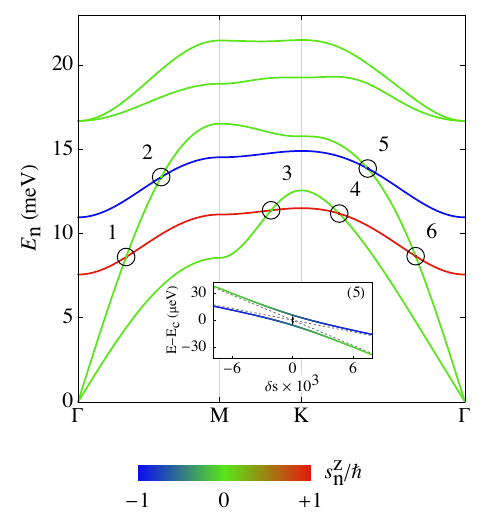}
\caption{Hybrid magnon--phonon spectrum along $\Gamma$--$M$--$K$--$\Gamma$, with $S=2$, $\tau=1/2$,
$(J,\lambda,D_o,D)=(1.57,11,0.7,0.1)$ meV,
$M=55.845\,\mathrm{u}$, $a=3.33419$~\AA, and
$\hbar^2(k_1,k_{2t},k_{2o})/M=(93,24.4,51.6)$ meV$^2$.
Color denotes the magnon spin of each hybrid mode,
$s_n^z/\hbar$. Circles 1--6 mark anticrossings.
Inset: crossing (5), solid and dashed
curves show coupled and uncoupled modes. The inset uses local coordinates
$\delta s=s-s_0$ and $E-E_c$, where $s_0$ is the position of minimum
separation between the two hybridized bands and $E_c$ is their mean
energy at $s_0$.
All six anticrossings are shown in the Supplemental Material.}
\label{fig:main-spectra}
\end{figure}

\begin{figure*}[!t]
\centering
\begin{minipage}[t]{0.538\linewidth}
\raggedright
\vspace{0pt}
\noindent(a)\par\smallskip
\includegraphics[width=\linewidth]{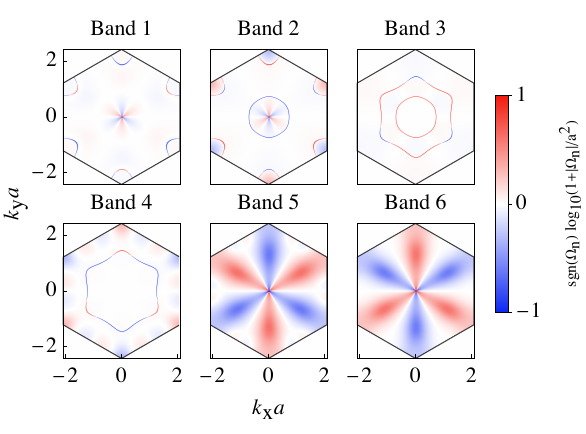}
\end{minipage}\hfill
\begin{minipage}[t]{0.441\linewidth}
\raggedright
\vspace{0pt}
\noindent(b)\par\smallskip
\vspace{0.0656\linewidth}
\includegraphics[width=\linewidth]{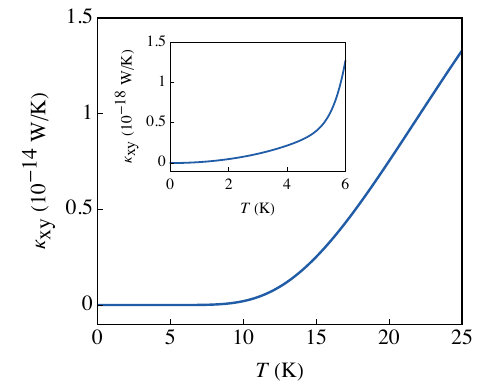}
\end{minipage}
\caption{(a) Berry curvature of the six positive-energy bands in the
first Brillouin zone, numbered in order of increasing energy. The parameters
are those of Fig.~\ref{fig:main-spectra}. 
(b) Thermal Hall conductivity, with the low-temperature
region enlarged in the inset.}
\label{fig:main-response}
\end{figure*}

\noindent\bluefour{\it Magnon--phonon excitations.}---Having established the
ferrimagnetic state, we turn to the magnetic
excitations and their coupling to lattice vibrations, which gives rise to nontrivial band topology. 
The in-plane
DM interaction mentioned above provides a microscopic
mechanism for this coupling~\cite{bao2023}. We retain the leading Heisenberg
part of $H_{ex}$ in $H_{min}$ and write the DM interaction as~\cite{ma,moriya1960}
\begin{equation}
H_D=D\sum_{\langle ij\rangle}
(\hat{\mathbf z}\times\hat{\mathbf e}_{ij})
\cdot(\mathbf S_i\times\mathbf S_j),
\label{eq:main-dm}
\end{equation}
where $\langle ij\rangle$ denotes nearest-neighbor bonds with $i\in t$
and $j\in o$, and $\hat{\mathbf e}_{ij}$ is the unit vector along the
instantaneous bond from $i$ to $j$. This weak DM interaction leaves the collinear order established by $H_{min}$ unchanged, while its modulation by ionic displacements generates magnon–phonon coupling~\cite{zhang2019mpi,park2020mpi} and its presence is further related to Berry curvature 
generation~\cite{Onose_2010,Zhang_2024}.
Keeping terms linear in both the ionic displacements $\udis_i$ and the
transverse spin fluctuations $\mathbf s_i=(S_i^x,S_i^y)^T$ gives~\cite{ma}
$H_{mp}^{(2)}=(DS/a)\sum_{\langle ij\rangle}
(\udis_j-\udis_i)^TQ_{ij}(\mathbf s_i+\mathbf s_j)$,
where $a$ is the equilibrium nearest-neighbor distance and
$Q_{ij}=\Id_2-\hat{\mathbf e}_{ij}^{\,0}
(\hat{\mathbf e}_{ij}^{\,0})^T$ projects the relative displacement
perpendicular to the equilibrium bond direction $\hat{\mathbf e}_{ij}^{\,0}$.

The lattice dynamics is described by a minimal model of the
in-plane ionic displacements,
\begin{equation}
H_p=\sum_i\frac{\mathbf P_i^2}{2M}
+\frac12\sum_{(ij)}k_{ij}
\bigl[\hat{\mathbf e}_{ij}^{\,0}\cdot(\udis_j-\udis_i)\bigr]^2,
\label{eq:main-phonon}
\end{equation}
where $\mathbf P_i$ is the momentum conjugate to $\udis_i$, and $M$ is the
Fe mass. The spring constants are $k_1$ for nearest-neighbor bonds
connecting the $t$ and $o$ sublattices and $k_{2t}$ ($k_{2o}$) for
next-nearest-neighbor bonds within the $t$ ($o$) sublattice. The latter are included to
ensure lattice rigidity~\cite{lubensky2015}. Together with the magnetic interactions
established above, the coupled system is described by
$H=H_{min}+H_p+H_D$.

In this model, the orbital pseudospin fluctuations decouple from
the magnon--phonon modes at harmonic order. We therefore work in the
ordered orbital sector $\tau_i^y=1/2$ and examine the coupled spin and
lattice fluctuations. A Holstein--Primakoff (HP) expansion~\cite{hp} about the
collinear state, followed by a bosonic Bogoliubov
transformation~\cite{colpa}, gives the six magnon--phonon branches shown
in Fig.~\ref{fig:main-spectra}. The hybridization opens gaps at the six marked
anticrossings. One is enlarged in the inset, and all six are shown in the
Supplemental Material.
The color denotes the magnon spin of each hybrid mode, $
s_n^z/\hbar
=\langle
-a_{\mathbf k}^{\dagger}a_{\mathbf k}
+b_{\mathbf k}^{\dagger}b_{\mathbf k}
\rangle_n$,
where $a$ and $b$ are the HP bosons for the tetrahedral and octahedral \(\text{Fe}^{2+}\) sublattices
respectively. Accordingly, the blue and red branches are dominated by spin fluctuations on the \(t\) and \(o\) sublattices respectively, whereas the green branches are predominantly phononic. The $t$-like magnon (blue)
branch lies higher in energy because of the strong SOC contribution.

The hybridization also gives rise to nontrivial band topology~\cite{go2019,zhang2020su3}. For a
positive-energy mode $v_n$ normalized by $v_n^\dagger g v_n=1$, the Berry
curvature is~\cite{shindou,ma}
$\Omega_n(\vk)=-2\operatorname{Im}
[(\partial_{k_x}v_n)^\dagger g\partial_{k_y}v_n]$,
where $g$ is the bosonic commutator metric.
As shown in Fig.~\ref{fig:main-response}(a), the Berry curvature is
concentrated near the anticrossings, where the hybridized eigenstates
vary rapidly with momentum. The two branches involved in an
anticrossing carry opposite signs of Berry curvature in its vicinity.
For an isolated band, integrating the curvature over the first Brillouin
zone (BZ) gives the Chern number,
$C_n=(2\pi)^{-1}\int_{\rm BZ}d^2k\,\Omega_n(\vk)$,
which can also be evaluated by the Fukui method~\cite{fukui}.
The two acoustic bands meet at $\Gamma$ and are therefore assigned a
joint Chern number $C_{1\oplus2}$. The remaining bands are isolated;
in particular, the coupling weakly splits the optical pair, allowing
$C_5$ and $C_6$ to be defined separately. For the parameters used in
Fig.~\ref{fig:main-spectra}, we obtain
$(C_{1\oplus2},C_3,C_4,C_5,C_6)=(-2,1,1,-1,1)$.

The Berry curvature of the hybrid magnon--phonon bands in
Fe$_2$Mo$_3$O$_8$ gives rise to a transverse heat current
under a longitudinal temperature gradient, providing an
experimentally accessible signature of the band geometry.
The corresponding thermal Hall conductivity is~\cite{matsumoto2011,matsumoto2014,ma}
\begin{equation}
\kappa_{xy}=-\frac{k_B^2T}{\hbar}\sum_{n=1}^{6}
\int_{\rm BZ}\frac{d^2k}{(2\pi)^2}\,
\mathcal W(\epsilon_{n\vk},T)\Omega_n(\vk),
\label{eq:main-thermal}
\end{equation}
where $\mathcal W(\epsilon,T)=c_2[n_B(\epsilon)]-\pi^2/3$,
$n_B$ is the Bose distribution, and
$c_2(x)=(1+x)[\ln(1+1/x)]^2-(\ln x)^2-2\operatorname{Li}_2(-x)$,
with $\operatorname{Li}_2$ the dilogarithm function.
Figure~\ref{fig:main-response}(b) shows the calculated thermal Hall
conductivity, which is positive at finite temperatures in the range up to 25 K. The inset resolves the low-temperature response.
These results demonstrate a finite zero-field thermal Hall response of the model.

\noindent\bluefour{\it Discussion.}---In summary, 
we have shown that the internal energy-level structure of $3d$ magnetic moments,
here realized through the inequivalent tetrahedral and octahedral sublattices 
of the $A_2$Mo$_3$O$_8$ honeycomb, constitutes a physical degree of freedom 
that is every bit as consequential as the spin itself. 
By resolving the atomic multiplet of the $A^{2+}$ ions and its interplay with SOC and exchange, 
we establish how these competing energy scales conspire to generate the emergent low-energy behaviour. 
The site-resolved theoretical framework established here for \(\text{Fe}_2\text{Mo}_3\text{O}_8\) 
provides a universal platform to rationalize the SOC in substituted polar \(M_2\text{Mo}_3\text{O}_8\) 
magnets. First, this model can be extended to the site-ordered \(\text{FeNiMo}_3\text{O}_8\) 
to isolate the tetrahedral \(\text{Fe}^{2+}\) contributions, 
thereby decoupling the complex interactions driving its high-temperature glassy state 
and exchange bias~\cite{yang2025}. Furthermore, this approach can be directly applied to the isostructural analogue 
\(\text{Ni}_2\text{Mo}_3\text{O}_8\) to evaluate how octahedral \(\text{Ni}^{2+}\) 
configurations alter the global magnetic anisotropy~\cite{morey2019,gao2023}. 
Finally, our framework remains highly relevant for the non-magnetic dilution limit in 
\(\text{FeZnMo}_3\text{O}_8\). By freezing out the tetrahedral magnetism with \(\text{Zn}^{2+}\)~\cite{streltsov2019}, 
the model can pristinely capture the unconstrained single-ion anisotropy and exchange pathways of the remaining octahedral \(\text{Fe}^{2+}\) sublattice, unveiling the mechanism 
behind its unique uniaxial ferromagnetic ground state~\cite{streltsov2019}.

In this perspective, our work underscores that the complex electronic multiplets of $3d$ transition-metal ions are not merely isolated atomic features, but active microscopic degrees of freedom that can fundamentally shape the emergent states of correlated Mott insulators. When embedded within the asymmetric sublattices of the $A_2\text{Mo}_3\text{O}_8$ honeycomb structure, these atomic multiplets intertwine with local crystal fields and exchange interactions, creating somewhat rich landscape of spin-orbit-entangled physics. The framework's capacity to systematically track these configurations, across the varying local environments presented by $\text{Fe}_2\text{Mo}_3\text{O}_8$, $\text{FeNiMo}_3\text{O}_8$, and $\text{FeZnMo}_3\text{O}_8$, reveals how subtle reshuffling of the electronic multiplets can vary between different ground states. These insights demonstrate that resolving the site-specific multiplet structure is essential for capturing the low-energy emergent behavior of complex oxides, highlighting the central role of spin-orbital-lattice entanglement in driving the rich emergent physics of multi-sublattice materials.

\noindent\bluefour{\it Acknowledgments.}---Hong-Hao Song thanks Bowen Ma for helpful discussions.
This work was supported 
by the NSFC (Grant No.~92565110 and 12574061), 
the Quantum Science and Technology--National Science and Technology Major Project 
(Grant No.~2025ZD0300500), 
and the BJNSF (Grant No.~F261004).

\bibliographystyle{apsrev4-2}
\let\originalbibfont\bibfont
\renewcommand{\bibfont}{\originalbibfont\setlength{\baselineskip}{0.98\baselineskip}}
\bibliography{Refs}
\let\bibfont\originalbibfont

\clearpage
\onecolumngrid

\begin{center}
\textbf{\large\resizebox{\textwidth}{!}{\shortstack{Supplemental Material for \\ ``Intertwined complex orbital order and ferrimagnetism in honeycomb molybdate Fe$_2$Mo$_3$O$_8$''}}}
\end{center}

\begingroup
\makeatletter

\let\appendix@toc@section\l@section
\let\l@section\@gobble@tw@
\let\l@title\@gobble@tw@
\let\l@subsection\@gobble@tw@
\let\l@subsubsection\@gobble@tw@
\let\l@paragraph\@gobble@tw@
\let\l@subparagraph\@gobble@tw@

\addtocontents{toc}{\protect\let\protect\l@section\protect\appendix@toc@section}
\tableofcontents
\makeatother
\endgroup

\renewcommand{\thefigure}{S\arabic{figure}}
\renewcommand{\theHfigure}{S\arabic{figure}}
\setcounter{figure}{0}
\renewcommand{\theequation}{S\arabic{equation}}
\renewcommand{\theHequation}{S\arabic{equation}}
\setcounter{equation}{0}
\renewcommand{\thesection}{\Roman{section}}
\setcounter{section}{0}
\setcounter{secnumdepth}{4}

\section{On the spin-orbit coupling of the tetrahedral site}

Here we give a detailed elaboration about the spin--orbit coupling (SOC) of the tetrahedral site. We focus on the 
$e_1$ manifold where there is a two-fold orbital degeneracy and the atomic SOC 
is active at the linear order. 
Since this manifold is more than half filled, the SOC is negative with
\begin{equation}
H_{soc} = -\lambda {\boldsymbol S}_{e_1} \cdot {\boldsymbol L},
\end{equation}
where ${\boldsymbol S}_{e_1}$ refers to the total spin from the $e_1$ manifold. 
As there is only one unpaired electron, we have $S_{e_1} =1/2$ here. 
After projecting to the $e_1$ manifold, the orbital angular momentum only remains
the $z$ component $L^z$. Using the pseudospin representation, $L^z \rightarrow 4\tau^y$. 

Now we include the Hund's coupling between the spin of the $e_1$ manifold and the 
spin of the half-filled upper $t$ manifold,
\begin{eqnarray}
H_{Hund} = -J_H {\boldsymbol S}_{e_1} \cdot {\boldsymbol S}_t,
\end{eqnarray}
where $S_t =3/2$. 
According to Hund's rule, ${\boldsymbol S}_{e_1}  $ and ${\boldsymbol S}_t$ are parallel locked,
and together form a total spin ${\boldsymbol S}$ with $S=2$. Thus, the effective 
SOC of the tetrahedral site becomes
\begin{equation}
H_{soc} = -\lambda {S}^z  {\tau}^y.
\end{equation}

\section{Construction of the quadratic Hamiltonian}
In this section, nearest-neighbor (NN) bonds $\langle ij\rangle$ in the main text are written
explicitly as $(i,i+\delta_\nu)$, where $i\in t$, $i+\delta_\nu\in o$,
and $\nu=1,2,3$ labels the three bond vectors defined below.
Likewise, next-nearest-neighbor (NNN) bonds are written as $(i,i+d_\mu)$,
with both sites on the same sublattice $\alpha=t,o$.
The three vectors $\mathbf d_\mu$ ($\mu=1,2,3$) are defined below.

\begin{figure}[tbp]
\centering
\includegraphics[width=.4\linewidth]{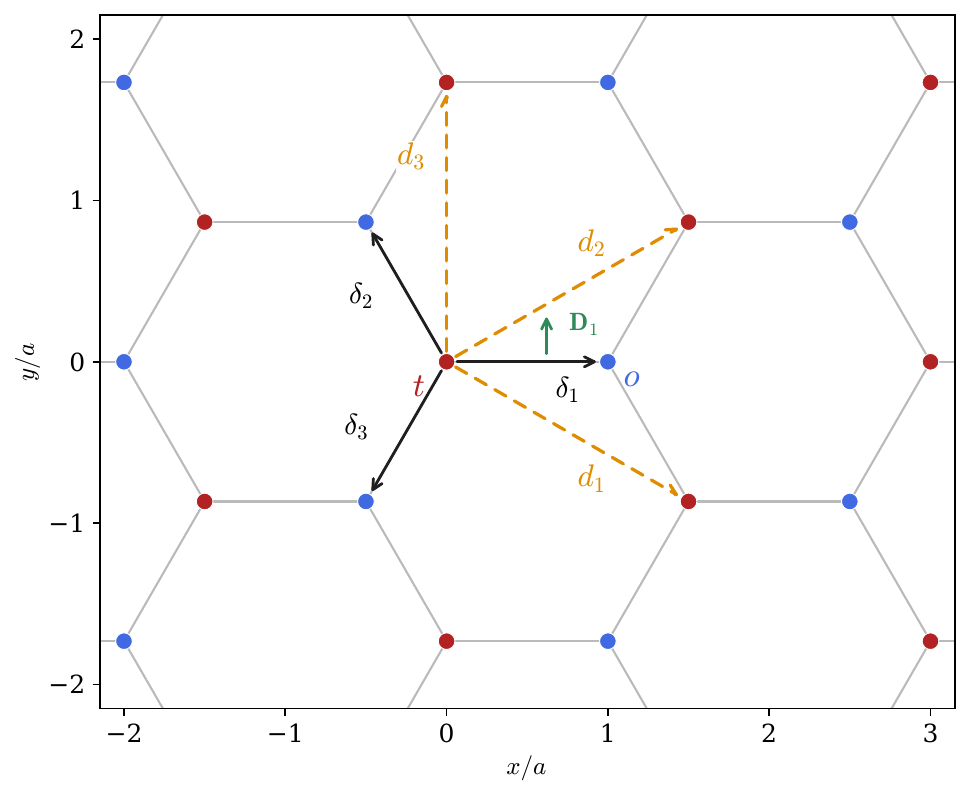}
\caption{Honeycomb geometry. Red and blue sites denote the $t$ and $o$
sublattices. Black arrows show the NN vectors
$\boldsymbol\delta_1$--$\boldsymbol\delta_3$; dashed orange arrows show
$\mathbf d_1$--$\mathbf d_3$ from a $t$ site to its same-sublattice neighbors.
The same NNN vectors apply to the $o$ sublattice. The green arrow indicates
$\mathbf D_1=D(\hat z\times\hat\delta_1)$.}
\label{fig:lattice}
\end{figure}

In this notation, the Hamiltonian studied in the main text reads
\begin{subequations}\label{eq:total-hamiltonian}
\begin{equation}
H=H_m+H_p+H_D,
\label{eq:total-real}
\end{equation}

\begin{equation}
H_m=-\lambda\sum_{i\in t}S_i^z\tau_i^y
-D_o\sum_{j\in o}(S_j^z)^2
+J\sum_{i\in t,\nu}\mathbf S_i\cdot\mathbf S_{i+\delta_\nu},
\label{eq:total-magnon}
\end{equation}

\begin{equation}
H_p={}\sum_i\frac{\mathbf P_i^2}{2M}
+\frac{k_1}{2}\sum_{i\in t,\nu}
[\hat\delta_\nu\cdot(\udis_{o,i+\delta_\nu}-\udis_{t,i})]^2+\frac12\sum_{\alpha=t,o}k_{2\alpha}\sum_{i\in\alpha,\mu=1}^3
[\hat d_\mu\cdot(\udis_{\alpha,i+d_\mu}-\udis_{\alpha,i})]^2,
\label{eq:total-phonon}
\end{equation}

\begin{equation}
H_D=D\sum_{i\in t,\nu}
(\hat{\mathbf z}\times\hat{\mathbf e}_{i,i+\delta_\nu})
\cdot(\mathbf S_i\times\mathbf S_{i+\delta_\nu}).
\label{eq:total-coupling}
\end{equation}
\end{subequations}
Here $t$ and $o$ denote the tetrahedral and octahedral sublattices respectively, and $a$ is the NN distance.
The NN bond vectors are $\boldsymbol\delta_1=a(1,0)$,
$\boldsymbol\delta_2=a(-1/2,\sqrt3/2)$, and
$\boldsymbol\delta_3=a(-1/2,-\sqrt3/2)$.
The three NNN vectors are
$\mathbf d_1=\boldsymbol\delta_1-\boldsymbol\delta_2$,
$\mathbf d_2=\boldsymbol\delta_1-\boldsymbol\delta_3$, and
$\mathbf d_3=\boldsymbol\delta_2-\boldsymbol\delta_3$.
The six NNNs of each site lie along $\pm\mathbf d_\mu$; summing over all
sites and only the three directions $\mathbf d_\mu$ counts each undirected
bond once.
We choose primitive vectors $\mathbf A_1=\mathbf d_1$ and
$\mathbf A_2=\mathbf d_2$, with intracell positions $r_t=0$ and
$r_o=\boldsymbol\delta_1$. The reciprocal vectors satisfy
$\mathbf B_i\cdot\mathbf A_j=2\pi\delta_{ij}$.
The elastic terms use the equilibrium unit vectors $\hat\delta_\nu$ and
$\hat d_\mu$, while $\hat{\mathbf e}_{ij}$ is the instantaneous bond
direction in the Dzyaloshinskii--Moriya interaction. The bond geometry is shown in
Fig.~\ref{fig:lattice}.

\subsection{Magnon Hamiltonian}
The Hamiltonian of the magnon sector is
\begin{equation}
H_m=-\lambda\sum_{i\in t}S_i^z\tau_i^y
-D_o\sum_{j\in o}(S_j^z)^2
+J\sum_{i\in t,\nu}\mathbf S_i\cdot\mathbf S_{i+\delta_\nu}.
\label{eq:H0}
\end{equation}
As discussed in the main text, the ground state is given by
\begin{equation}
S=2,\quad T=\tfrac12,\qquad
\langle\mathbf S_t\rangle=S\hat{\mathbf z},\quad
\langle\mathbf S_o\rangle=-S\hat{\mathbf z},\quad
\langle\boldsymbol\tau_t\rangle=T\hat{\mathbf y}.
\label{eq:reference}
\end{equation}

Using the Holstein--Primakoff (HP) transformations~\cite{hp} 
\begin{subequations}\label{eq:HP}
\begin{equation}
S_{t,i}^{+}=\sqrt{2S-n_{a,i}}\,a_i\simeq\sqrt{2S}\,a_i,
\qquad S_{t,i}^{-}=a_i^\dagger\sqrt{2S-n_{a,i}}\simeq\sqrt{2S}\,a_i^\dagger,
\qquad S_{t,i}^{z}=S-n_{a,i},
\label{eq:HP-t}
\end{equation}

\begin{equation}
S_{o,j}^{+}=b_j^\dagger\sqrt{2S-n_{b,j}}\simeq\sqrt{2S}\,b_j^\dagger,
\qquad S_{o,j}^{-}=\sqrt{2S-n_{b,j}}\,b_j\simeq\sqrt{2S}\,b_j,
\qquad S_{o,j}^{z}=-S+n_{b,j},
\label{eq:HP-o}
\end{equation}

\begin{equation}
\tau_i^{+_y}=\sqrt{2T-n_{c,i}}\,c_i\simeq\sqrt{2T}\,c_i,
\qquad \tau_i^{-_y}=c_i^\dagger\sqrt{2T-n_{c,i}}\simeq\sqrt{2T}\,c_i^\dagger,
\qquad \tau_i^y=T-n_{c,i},
\label{eq:HP-tau}
\end{equation}
\end{subequations}
the quadratic magnetic Hamiltonian becomes
\begin{subequations}\label{eq:magnetic-quadratic}
\begin{equation}
H_m^{(2)}=E_{\rm cl}+A_t\sum_i n_{a,i}+A_o\sum_j n_{b,j}
+\lambda S\sum_{i\in t}n_{c,i}
+JS\sum_{i,\nu}(a_i b_{i+\delta_\nu}+a_i^\dagger b_{i+\delta_\nu}^\dagger),
\label{eq:realquad}
\end{equation}
\begin{equation}
A_t=3JS+\lambda T,\quad A_o=3JS+(2S-1)D_o, \quad 
E_{\rm cl}/N={}-3JS^2-\lambda ST-D_oS^2.
\end{equation}
\end{subequations}
Here $n_{a,i}=a_i^\dagger a_i$, $n_{b,j}=b_j^\dagger b_j$, and
$n_{c,i}=c_i^\dagger c_i$, while $N$ is the number of sites on either
sublattice.
Although exchange couples the two sublattices, the two bare magnon
branches retain predominantly $t$ and $o$ character. For the parameters
used in the main text, spin fluctuations on the $t$ and $o$ sublattices
have characteristic energy scales of $A_t\simeq14.9$ meV and
$A_o\simeq11.5$ meV, respectively.
Since $[H,\tau_i^y]=0$ and the pseudospin excitation is decoupled from
the magnon--phonon modes at harmonic order, we work in the conserved
sector $\tau_i^y=T$, or equivalently $n_{c,i}=0$.
In the subsequent analysis, we therefore retain only the
$a$ and $b$ bosons in the magnetic sector.

To express the resulting magnon Hamiltonian in momentum space, we use the
Fourier transformation
\begin{equation}
a_i=\frac1{\sqrt N}\sum_\vk e^{i\vk\cdot\vR_i}a_\vk,
\qquad b_j=\frac1{\sqrt N}\sum_\vk e^{i\vk\cdot\vR_j}b_\vk.
\label{eq:fourier}
\end{equation}
Here $\vk$ runs over the full Brillouin zone (BZ), and $\vR_i$ and $\vR_j$
denote the positions of the $t$ and $o$ sites.
With the Nambu basis $\Psi_\vk=(a_\vk,b_\vk,a_{-\vk}^\dagger,b_{-\vk}^\dagger)^T$, the quadratic magnon Hamiltonian then takes the matrix form
\begin{equation}
H_m^{(2)}=\frac12\sum_\vk\Psi_\vk^\dagger h_m(\vk)\Psi_\vk+\mathrm{const.},
\qquad
h_m=\begin{pmatrix}
A_t&0&0&C_\vk\\
0&A_o&C_\vk^*&0\\
0&C_\vk&A_t&0\\
C_\vk^*&0&0&A_o
\end{pmatrix}.
\label{eq:hm}
\end{equation}
Here $C_\vk=JS\gamma_\vk$,
and $\gamma_\vk=\sum_{\nu=1}^3e^{i\vk\cdot\boldsymbol\delta_\nu}$ is the
NN structure factor.

\subsection{Phonon Hamiltonian}
The Hamiltonian of the phonon sector is
\begin{equation*}
H_p={}\sum_i\frac{\mathbf P_i^2}{2M}
+\frac{k_1}{2}\sum_{i\in t,\nu}
[\hat\delta_\nu\cdot(\udis_{o,i+\delta_\nu}-\udis_{t,i})]^2+\frac12\sum_{\alpha=t,o}k_{2\alpha}\sum_{i\in\alpha,\mu=1}^3
[\hat d_\mu\cdot(\udis_{\alpha,i+d_\mu}-\udis_{\alpha,i})]^2.
\end{equation*}
We introduce the dimensionless displacement and momentum
\begin{equation}
\utv_i=\frac{\udis_i}{a},\qquad
\ptv_i=\frac{a\mathbf P_i}{\hbar}.
\label{eq:scaling}
\end{equation}
They satisfy the canonical commutation relations
$
[\widetilde u_{i\mu},\widetilde p_{j\eta}]
=i\delta_{ij}\delta_{\mu\eta}$,
where $\mu,\eta\in\{x,y\}$.
Then the phonon Hamiltonian can be written as
\begin{equation}
H_p={}\frac{E_0}{2}\sum_i\ptv_i^2
+\frac{w_1^2}{2E_0}\sum_{i\in t,\nu}
\bigl[\hat\delta_\nu\cdot(\utv_{o,i+\delta_\nu}-\utv_{t,i})\bigr]^2
+\frac{1}{2E_0}\sum_{\alpha=t,o}w_{2\alpha}^2
\sum_{i\in\alpha,\mu=1}^3
\bigl[\hat d_\mu\cdot(\utv_{\alpha,i+d_\mu}-\utv_{\alpha,i})\bigr]^2,
\label{eq:phonon-dimensionless}
\end{equation}
where $E_0=\hbar^2/(Ma^2)$ and $w_\ell^2=\hbar^2k_\ell/M$
for $\ell=1,2t,2o$.
Introducing the longitudinal projectors
$P_\nu=\hat\delta_\nu\hat\delta_\nu^T$ and
$P'_\mu=\hat d_\mu\hat d_\mu^T$, we can equivalently write
\begin{equation}
\begin{aligned}
H_p={}&\frac{E_0}{2}\sum_i\ptv_i^T\ptv_i+\frac{w_1^2}{2E_0}\sum_{i\in t,\nu}
(\utv_{o,i+\delta_\nu}-\utv_{t,i})^T
P_\nu(\utv_{o,i+\delta_\nu}-\utv_{t,i})\\
&+\frac{1}{2E_0}\sum_{\alpha=t,o}w_{2\alpha}^2
\sum_{i\in\alpha,\mu=1}^3
(\utv_{\alpha,i+d_\mu}-\utv_{\alpha,i})^T
P'_\mu(\utv_{\alpha,i+d_\mu}-\utv_{\alpha,i}).
\end{aligned}
\label{eq:phonon-projectors}
\end{equation}
The displacement and momentum vectors are Fourier transformed as
\begin{equation}
\utv_{\alpha,i}=\frac{1}{\sqrt N}\sum_\vk
e^{i\vk\cdot\vR_{\alpha,i}}\utv_{\alpha,\vk},\qquad
\ptv_{\alpha,i}=\frac{1}{\sqrt N}\sum_\vk
e^{i\vk\cdot\vR_{\alpha,i}}\ptv_{\alpha,\vk},
\label{eq:phonon-fourier}
\end{equation}
where $\alpha=t,o$ and $\vR_{\alpha,i}$ is the actual site position;
each vector contains the $x$ and $y$ components.
Hermiticity implies
$
\utv_{\alpha,\vk}^{\dagger}=\utv_{\alpha,-\vk}^{ T},
\ptv_{\alpha,\vk}^{\dagger}=\ptv_{\alpha,-\vk}^{ T}.
$
Collecting the components into the column vectors
\[
\utv_\vk=\bigl(\widetilde u_{t,x}(\vk),\widetilde u_{t,y}(\vk),
\widetilde u_{o,x}(\vk),\widetilde u_{o,y}(\vk)\bigr)^T,
\]
\[
\ptv_\vk=\bigl(\widetilde p_{t,x}(\vk),\widetilde p_{t,y}(\vk),
\widetilde p_{o,x}(\vk),\widetilde p_{o,y}(\vk)\bigr)^T,
\]
the Hamiltonian becomes
\begin{equation}
H_p=\frac12\sum_\vk\left[
\utv_\vk^\dagger\frac{\Phi(\vk)}{E_0}\utv_\vk
+\ptv_\vk^\dagger E_0\Id_4\ptv_\vk\right].
\label{eq:phonon-H}
\end{equation}
where the elastic matrix $\Phi(\vk)$ is
\begin{subequations}\label{eq:phi}
\begin{equation}
\Phi(\vk)=\begin{pmatrix}A_{p,t}&B_p\\B_p^\dagger&A_{p,o}\end{pmatrix},
\end{equation}
\begin{equation}
A_{p,\alpha}=\frac32w_1^2\Id_2
+2w_{2\alpha}^2\sum_{\mu=1}^3[1-\cos(\vk\cdot\mathbf d_\mu)]P'_\mu,
\qquad
B_p=-w_1^2\sum_{\nu=1}^3e^{i\vk\cdot\boldsymbol\delta_\nu}P_\nu.
\end{equation}
\end{subequations}
With $\xi=3ak_x/2$ and $\zeta=\sqrt3ak_y/2$, the explicit blocks are
($\alpha=t,o$)
\begin{subequations}\label{eq:phi-explicit}
\begin{align}
A_{p,\alpha}&=\begin{pmatrix}
\frac32w_1^2+3w_{2\alpha}^2(1-\cos\xi\cos\zeta)
&\sqrt3w_{2\alpha}^2\sin\xi\sin\zeta\\
\sqrt3w_{2\alpha}^2\sin\xi\sin\zeta
&\frac32w_1^2+w_{2\alpha}^2[3-\cos\xi\cos\zeta-2\cos(2\zeta)]
\end{pmatrix},\\
B_p&=-w_1^2\begin{pmatrix}
e^{iak_x}+\frac12e^{-iak_x/2}\cos\zeta&-\frac{i\sqrt3}{2}e^{-iak_x/2}\sin\zeta\\
-\frac{i\sqrt3}{2}e^{-iak_x/2}\sin\zeta&\frac32e^{-iak_x/2}\cos\zeta
\end{pmatrix}.
\end{align}
\end{subequations}

\subsection{Magnon--phonon coupling}
The magnon--phonon coupling sector is
\begin{equation}
H_D=D\sum_{i\in t,\nu}
(\hat{\mathbf z}\times\hat{\mathbf e}_{i,i+\delta_\nu})
\cdot(\mathbf S_i\times\mathbf S_{i+\delta_\nu}).
\label{eq:DM}
\end{equation}
For a bond directed from $i\in t$ to $j=i+\delta_\nu\in o$, we use
$(\mathbf A\times\mathbf B)\cdot(\mathbf C\times\mathbf D)
=(\mathbf A\cdot\mathbf C)(\mathbf B\cdot\mathbf D)
-(\mathbf A\cdot\mathbf D)(\mathbf B\cdot\mathbf C)$
to rewrite the bond energy as
\begin{equation}
H_{D,ij}=D\left[S_i^z\,\hat{\mathbf e}_{ij}\cdot\mathbf S_j
-S_j^z\,\hat{\mathbf e}_{ij}\cdot\mathbf S_i\right].
\end{equation}
Since the bond lies in the plane, only the in-plane spin components
enter these dot products. Defining $\mathbf s_i=(S_i^x,S_i^y)^T$ and
$\mathbf s_j=(S_j^x,S_j^y)^T$, and substituting
$S_i^z=S-n_{a,i}$ and $S_j^z=-S+n_{b,j}$, we obtain
\begin{equation}\label{eq:DM-spin-expansion}
H_{D,ij}=DS\,\hat{\mathbf e}_{ij}\cdot(\mathbf s_i+\mathbf s_j)
-D\,\hat{\mathbf e}_{ij}\cdot
(n_{a,i}\mathbf s_j+n_{b,j}\mathbf s_i)
\simeq DS\,\hat{\mathbf e}_{ij}\cdot(\mathbf s_i+\mathbf s_j).
\end{equation}
At quadratic order, we retain only the first term.

Here $\hat{\mathbf e}_{ij}=\mathbf R_{ij}/|\mathbf R_{ij}|$ is the
unit vector along the instantaneous bond directed from site $i$ to site $j$.
Let $\Delta\udis_{ij}=\udis_j-\udis_i$ and
$\mathbf R_{ij}=a\hat\delta_\nu+\Delta\udis_{ij}$.
To first order in displacement,
\begin{subequations}\label{eq:bond-direction-expansion}
\begin{align}
|\mathbf R_{ij}|
&=a\sqrt{1+\frac{2\hat\delta_\nu\cdot\Delta\udis_{ij}}{a}
+\frac{|\Delta\udis_{ij}|^2}{a^2}}
=a+\hat\delta_\nu\cdot\Delta\udis_{ij}+O(u^2/a),\\
\frac1{|\mathbf R_{ij}|}
&=\frac1a\left[1-\frac{\hat\delta_\nu\cdot\Delta\udis_{ij}}{a}
+O(u^2/a^2)\right],\\
\hat{\mathbf e}_{ij}
&=(a\hat\delta_\nu+\Delta\udis_{ij})\frac1{|\mathbf R_{ij}|}
\nonumber\\
&=\hat\delta_\nu+\frac1a\left[\Delta\udis_{ij}
-\hat\delta_\nu(\hat\delta_\nu\cdot\Delta\udis_{ij})\right]
+O(u^2/a^2)\nonumber\\
&=\hat\delta_\nu+\frac1aQ_\nu\Delta\udis_{ij}+O(u^2/a^2),
\qquad Q_\nu=\Id_2-P_\nu=\Id_2-\hat\delta_\nu\hat\delta_\nu^T.
\end{align}
\end{subequations}

Combining Eqs.~\eqref{eq:DM-spin-expansion} and
\eqref{eq:bond-direction-expansion} in Eq.~\eqref{eq:DM}, and retaining
terms of total degree at most two in spin and displacement fluctuations, gives
\begin{equation}\label{eq:mp-real-derivation}
H_D \simeq DS\sum_{i\in t,\nu}\hat\delta_\nu\cdot
(\mathbf s_{t,i}+\mathbf s_{o,i+\delta_\nu})+\frac{DS}{a}\sum_{i\in t,\nu}
(\udis_{o,i+\delta_\nu}-\udis_{t,i})^TQ_\nu
(\mathbf s_{t,i}+\mathbf s_{o,i+\delta_\nu}).
\end{equation}
The first term vanishes upon summing over bonds because
$\sum_\nu\hat\delta_\nu=0$, leaving the quadratic magnon--phonon coupling
\begin{equation}
H_{mp}^{(2)}=DS\sum_{i\in t,\nu}
(\utv_{o,i+\delta_\nu}-\utv_{t,i})^TQ_\nu
(\mathbf s_{t,i}+\mathbf s_{o,i+\delta_\nu}).
\label{eq:mp-real}
\end{equation}

The Fourier transform of the in-plane spins follows from the boson transforms in Eq.~\eqref{eq:fourier}. 
Substituting Eq.~\eqref{eq:fourier} into the
linear HP relations in Eqs.~\eqref{eq:HP-t} and \eqref{eq:HP-o} yields
\begin{subequations}\label{eq:transverse-hp}
\begin{align}
\mathbf s_{t,i}
&\simeq\sqrt{S/2}\begin{pmatrix}a_i+a_i^\dagger\\-i(a_i-a_i^\dagger)\end{pmatrix}
=\frac1{\sqrt N}\sum_\vk e^{i\vk\cdot\vR_i}
\sqrt{S/2}\begin{pmatrix}a_\vk+a_{-\vk}^\dagger\\-i(a_\vk-a_{-\vk}^\dagger)\end{pmatrix},\\
\mathbf s_{o,j}
&\simeq\sqrt{S/2}\begin{pmatrix}b_j+b_j^\dagger\\+i(b_j-b_j^\dagger)\end{pmatrix}
=\frac1{\sqrt N}\sum_\vk e^{i\vk\cdot\vR_j}
\sqrt{S/2}\begin{pmatrix}b_\vk+b_{-\vk}^\dagger\\+i(b_\vk-b_{-\vk}^\dagger)\end{pmatrix}.
\end{align}
\end{subequations}
These relations can be rewritten as
\begin{equation}
\mathbf s_{\alpha,i}=\frac1{\sqrt N}\sum_\vk
 e^{i\vk\cdot\vR_{\alpha,i}}\mathbf s_{\alpha,\vk},
\qquad \alpha=t,o,
\label{eq:spin-fourier}
\end{equation}
with $\mathbf s_{t,\vk}=\sqrt{S/2}(a_\vk+a_{-\vk}^\dagger,
-i(a_\vk-a_{-\vk}^\dagger))^T$ and
$\mathbf s_{o,\vk}=\sqrt{S/2}(b_\vk+b_{-\vk}^\dagger,
+i(b_\vk-b_{-\vk}^\dagger))^T$.

Using the Fourier transforms of $\utv$ and $\mathbf s$ in
Eqs.~\eqref{eq:phonon-fourier} and \eqref{eq:spin-fourier}, the four products
for a fixed bond direction $\nu$ become
\begin{subequations}\label{eq:four-bond-products}
\begin{align}
-\sum_i\utv_{t,i}^TQ_\nu\mathbf s_{t,i}
&=-\sum_\vk\utv_{t,\vk}^\dagger Q_\nu\mathbf s_{t,\vk},\\
-\sum_i\utv_{t,i}^TQ_\nu\mathbf s_{o,i+\delta_\nu}
&=-\sum_\vk e^{i\vk\cdot\boldsymbol\delta_\nu}
\utv_{t,\vk}^\dagger Q_\nu\mathbf s_{o,\vk},\\
\sum_i\utv_{o,i+\delta_\nu}^TQ_\nu\mathbf s_{t,i}
&=\sum_\vk e^{-i\vk\cdot\boldsymbol\delta_\nu}
\utv_{o,\vk}^\dagger Q_\nu\mathbf s_{t,\vk},\\
\sum_i\utv_{o,i+\delta_\nu}^TQ_\nu\mathbf s_{o,i+\delta_\nu}
&=\sum_\vk\utv_{o,\vk}^\dagger Q_\nu\mathbf s_{o,\vk}.
\end{align}
\end{subequations}
With
$\mathbf s_\vk=(s_{t,x}(\vk),s_{t,y}(\vk),s_{o,x}(\vk),s_{o,y}(\vk))^T$,
these terms give
\begin{subequations}\label{eq:displacement-spin-matrix}
\begin{align}
H_{mp}^{(2)}&=DS\sum_\vk\utv_\vk^\dagger\mathsf T(\vk)\mathbf s_\vk,\\
\mathsf T(\vk)&=\sum_\nu\begin{pmatrix}
-Q_\nu&-e^{i\vk\cdot\boldsymbol\delta_\nu}Q_\nu\\
e^{-i\vk\cdot\boldsymbol\delta_\nu}Q_\nu&Q_\nu
\end{pmatrix}.
\end{align}
\end{subequations}
By the definition of $\mathbf s_\vk$ and
$\Psi_\vk=(a_\vk,b_\vk,a_{-\vk}^\dagger,b_{-\vk}^\dagger)^T$,
these relations become
\begin{equation}
\mathbf s_\vk=\sqrt{S/2}\,L\Psi_\vk,\qquad
L=\begin{pmatrix}
1&0&1&0\\-i&0&i&0\\0&1&0&1\\0&i&0&-i
\end{pmatrix}.
\label{eq:spin-HP-matrix}
\end{equation}
Substituting this relation into Eq.~\eqref{eq:displacement-spin-matrix} gives:
\begin{subequations}\label{eq:coupling-final}
\begin{align}
H_{mp}^{(2)}&=\sum_\vk\utv_\vk^\dagger F(\vk)\Psi_\vk
=\frac12\sum_\vk
\left[\utv_\vk^\dagger F(\vk)\Psi_\vk+
\Psi_\vk^\dagger F(\vk)^\dagger\utv_\vk\right],\\
F(\vk)&=DS\sqrt{S/2}\,\mathsf T(\vk)L.\label{eq:F}
\end{align}
\end{subequations}

\section{Bosonic diagonalization}
In the twelve-component basis $X_\vk=(\Psi_\vk^T,\utv_\vk^T,\ptv_\vk^T)^T$,
the Hamiltonian can be written as
\begin{equation}
H^{(2)}=\frac12\sum_\vk X_\vk^\dagger\mathcal H(\vk)X_\vk+\mathrm{const},
\qquad
\mathcal H=\begin{pmatrix}
h_m&F^\dagger&0\\F&\Phi/E_0&0\\0&0&E_0\Id_4
\end{pmatrix}.
\label{eq:fullH}
\end{equation}
The commutators of the mixed spin--displacement--momentum basis give
$[X_{\vk,i},X_{\vk',j}^\dagger]=g_{ij}\delta_{\vk,\vk'}$, with
\begin{equation}
g=\begin{pmatrix}\Sigma_z&0&0\\0&0&i\Id_4\\0&-i\Id_4&0\end{pmatrix},
\qquad \Sigma_z=\operatorname{diag}(1,1,-1,-1),\qquad g^2=\Id_{12}.
\end{equation}
The Heisenberg equation is
\[
i\hbar\frac{dX_\vk}{dt}
=[X_\vk,H^{(2)}]
=g\mathcal H(\vk)X_\vk.
\]
Substitution of $X_\vk\propto v_n e^{-i\epsilon_n t/\hbar}$ gives the
dynamical eigenproblem rather than an ordinary diagonalization of
$\mathcal H$. The normal modes solve
\begin{equation}
g\mathcal H v_n=\epsilon_n v_n,\qquad
v_n^\dagger gv_m=s_n\delta_{nm},\qquad s_n=\operatorname{sgn}\epsilon_n.
\label{eq:eigenproblem}
\end{equation}

We use the Colpa method~\cite{colpa} to solve the eigenproblem.
A positive-definite $\mathcal H$ can be factored as
\begin{equation}
\mathcal H=\mathsf R\mathsf R^\dagger.
\label{eq:colpa-construction}
\end{equation}
Here $\mathsf R$ is a lower-triangular Cholesky factor. Then we construct the Hermitian matrix
\begin{equation}
\mathsf K=\mathsf R^\dagger g\mathsf R=\mathsf K^\dagger,
\end{equation}
which can be diagonalized with orthonormal eigenvectors:
\begin{equation}
\mathsf K u_n=\epsilon_n u_n,\qquad u_n^\dagger u_m=\delta_{nm}.
\end{equation}
Since $\mathsf R^\dagger(g\mathcal H)(\mathsf R^\dagger)^{-1}=\mathsf K$,
the original dynamical problem has the same real eigenvalues. Set
\begin{equation}
v_n=(\mathsf R^\dagger)^{-1}u_n\sqrt{|\epsilon_n|}.
\end{equation}
This gives $g\mathcal H v_n=\epsilon_n v_n$. To check its normalization, use
\begin{equation}
\mathsf K^{-1}=\mathsf R^{-1}g(\mathsf R^\dagger)^{-1}
\end{equation}
to obtain
\begin{equation}
v_n^\dagger gv_m
=\sqrt{|\epsilon_n\epsilon_m|}\,u_n^\dagger\mathsf K^{-1}u_m
=s_n\delta_{nm},\qquad s_n=\operatorname{sgn}\epsilon_n.
\label{eq:colpa-normalization}
\end{equation}

Collecting all twelve eigenvectors into $V=(v_1,\ldots,v_{12})$ and
writing $\Sigma=\operatorname{diag}(s_1,\ldots,s_{12})$, the normalization
condition becomes
\begin{equation}
V^\dagger gV=\Sigma.
\end{equation}
Since $\Sigma^2=\Id_{12}$, multiplying by $\Sigma$ gives a left inverse,
and therefore
\begin{equation}
V^{-1}=\Sigma V^\dagger g.
\end{equation}
Substituting this into $VV^{-1}=\Id_{12}$ yields the completeness relation
\begin{equation}
\sum_{n=1}^{12}s_n v_n v_n^\dagger g=\Id_{12}.
\label{eq:colpa-completeness}
\end{equation}

\section{Magnon spin}
\label{sec:angular-momentum}
The HP and Fourier transformations give the magnon spin operator~\cite{ma}
\begin{equation}
\frac{\widehat S_z}{\hbar}
=-\sum_{i\in t}a_i^\dagger a_i+\sum_{j\in o}b_j^\dagger b_j
=\sum_\vk(-a_\vk^\dagger a_\vk+b_\vk^\dagger b_\vk).
\end{equation}
In the mixed basis this becomes
\begin{equation}
\frac{\widehat S_z}{\hbar}
=\frac12\sum_\vk X_\vk^\dagger\mathcal S X_\vk,
\label{eq:spin-kernel}
\end{equation}
with
\begin{equation}
\mathcal S=\operatorname{diag}(M_s,0_4,0_4),\qquad
M_s=\operatorname{diag}(-1,1,-1,1).
\end{equation}
The constants from the two sublattices cancel.

For a positive-energy mode $(n,\vk)$, write the normalized eigenvector as
\begin{equation}
v_{n\vk}=(\alpha_{n\vk},\beta_{n\vk},\bar\alpha_{n\vk},
\bar\beta_{n\vk},\mathbf u_{n\vk}^T,\mathbf p_{n\vk}^T)^T,\qquad
v_{n\vk}^\dagger gv_{n\vk}=1.
\label{eq:positive-mode-components}
\end{equation}
Here $n=1,\ldots,6$ labels the positive-energy branches. The bars label
creation-sector amplitudes, not complex conjugation; $\mathbf u_{n\vk}$
and $\mathbf p_{n\vk}$ contain the dimensionless displacement and momentum
components. For this observable, the annihilation and creation terms
contribute equally per excitation, canceling the prefactor $1/2$
in the quadratic form. The magnon spin per excitation, relative
to the Bogoliubov vacuum, is
\begin{equation}
\frac{S_{z,n}}{\hbar}=v_n^\dagger\mathcal S v_n
=-|\alpha_n|^2+|\beta_n|^2-|\bar\alpha_n|^2+|\bar\beta_n|^2,
\end{equation}
where the momentum label has been suppressed.
The spectrum is colored by $s_n^z/\hbar$, with $s_n^z=S_{z,n}$.

\section{Berry curvature}
For an isolated positive-energy band, the metric Berry connection and
curvature are~\cite{shindou,ma}
\begin{equation}
\mathcal A_{n,\mu}=i v_n^\dagger g\partial_{k_\mu}v_n,
\qquad
\Omega_n=\partial_{k_x}\mathcal A_{n,y}-\partial_{k_y}\mathcal A_{n,x}.
\end{equation}
Since $g$ is momentum independent, taking the curl gives
\begin{equation}
\Omega_n=-2\operatorname{Im}
[(\partial_{k_x}v_n)^\dagger g\partial_{k_y}v_n].
\label{eq:berry-derivative}
\end{equation}
Inserting $\Id=\sum_m s_m v_m v_m^\dagger g$ into
Eq.~\eqref{eq:berry-derivative} gives
\begin{equation}
\Omega_n=-2\operatorname{Im}\sum_{m=1}^{12}s_m
\bigl[(\partial_{k_x}v_n)^\dagger gv_m\bigr]
\bigl[v_m^\dagger g\partial_{k_y}v_n\bigr].
\label{eq:berry-completeness}
\end{equation}
The $m=n$ term is real and does not contribute.
To evaluate the remaining overlaps, differentiate
$\mathcal H v_n=\epsilon_ngv_n$:
\begin{equation}
(\partial_{k_\mu}\mathcal H)v_n+\mathcal H\partial_{k_\mu}v_n
=(\partial_{k_\mu}\epsilon_n)gv_n
+\epsilon_ng\partial_{k_\mu}v_n.
\end{equation}
Multiplying by $v_m^\dagger$ and using
$v_m^\dagger\mathcal H=\epsilon_mv_m^\dagger g$ yields
\begin{equation}
v_m^\dagger(\partial_{k_\mu}\mathcal H)v_n
=(\partial_{k_\mu}\epsilon_n)v_m^\dagger gv_n
+(\epsilon_n-\epsilon_m)v_m^\dagger g\partial_{k_\mu}v_n.
\end{equation}
For $m\ne n$, metric orthogonality removes the first term. Thus, for
$\epsilon_n\ne\epsilon_m$,
\begin{equation}
v_m^\dagger g\partial_{k_\mu}v_n
=\frac{v_m^\dagger(\partial_{k_\mu}\mathcal H)v_n}
{\epsilon_n-\epsilon_m}.
\label{eq:derivative-overlap}
\end{equation}
Substituting Eq.~\eqref{eq:derivative-overlap} and its complex conjugate
into Eq.~\eqref{eq:berry-completeness} yields
\begin{equation}
\Omega_n(\vk)=-2\operatorname{Im}
\sum_{\substack{m=1\\m\ne n}}^{12}s_m
\frac{(v_n^\dagger\partial_{k_x}\mathcal H v_m)
(v_m^\dagger\partial_{k_y}\mathcal H v_n)}
{(\epsilon_n-\epsilon_m)^2}.
\label{eq:kubo}
\end{equation}

\section{Chern numbers}
For an isolated positive band,
\begin{equation}
C_n=\frac1{2\pi}\int_{\rm BZ}\Omega_n(\vk)\,d^2k.
\end{equation}
This integral defines the invariant. Numerically, we compute it using
the Fukui--Hatsugai--Suzuki lattice construction~\cite{fukui}, directly
from metric eigenvector overlaps on a reciprocal-cell mesh, without
numerically integrating the sampled Berry curvature.
For neighboring vertices $i,j$, define
\begin{equation}
U_{ij}^{(n)}=\frac{v_n(\vk_i)^\dagger gv_n(\vk_j)}
{|v_n(\vk_i)^\dagger gv_n(\vk_j)|}.
\end{equation}

For a short link, metric normalization gives
\begin{equation}
v_n(\vk)^\dagger gv_n(\vk+d\vk)
=1-i\mathcal A_n\cdot d\vk+O(|d\vk|^2).
\end{equation}
Normalization retains only the phase:
\begin{equation}
U_{ij}^{(n)}
=\exp[-i\mathcal A_n(\vk_i)\cdot(\vk_j-\vk_i)]
+O(|\vk_j-\vk_i|^2).
\end{equation}
Multiplying the links along a small counterclockwise plaquette
$(0,1,2,3)$ adds their phases. In the continuum limit, the accumulated
phase is the line integral of the Berry connection:
\begin{equation}
U_{01}^{(n)}U_{12}^{(n)}U_{23}^{(n)}U_{30}^{(n)}
\simeq\exp\left[-i\oint_{\partial\square}\mathcal A_n\cdot d\vk\right].
\end{equation}
Stokes' theorem relates this line integral to the enclosed Berry curvature,
\begin{equation}
\oint_{\partial\square}\mathcal A_n\cdot d\vk
=\int_\square\Omega_n(\vk)\,d^2k.
\end{equation}
Thus the negative phase of the link product gives the Berry flux through
the plaquette:
\begin{equation}
\mathcal F_{n,\square}
=-\arg(U_{01}^{(n)}U_{12}^{(n)}U_{23}^{(n)}U_{30}^{(n)}).
\label{eq:plaquette-flux}
\end{equation}
Here $\arg$ denotes the principal phase; the continuum flux is recovered
when the mesh resolves each plaquette flux within $(-\pi,\pi)$.
Under $v_i\to e^{i\chi_i}v_i$, the link transforms as
$U_{ij}^{(n)}\to e^{i(\chi_j-\chi_i)}U_{ij}^{(n)}$.
The phases cancel around the closed loop, making both the plaquette
flux and the discrete Chern number independent of the eigenvector phases.
Summing over the plaquettes covering the BZ gives
\begin{equation}
C_n^{\rm mesh}=\frac1{2\pi}\sum_\square\mathcal F_{n,\square}.
\label{eq:FHS}
\end{equation}

When bands become
degenerate, the Chern number is defined for their joint subspace, provided
it remains separated from the other bands throughout the BZ.
For a group $\mathcal G$ of $r$ positive-energy bands, let
$V_{\mathcal G}=(v_{n_1},\ldots,v_{n_r})$. The corresponding link is
\begin{equation}
U_{ij}^{\mathcal G}=
\frac{\det[V_{\mathcal G}(\vk_i)^\dagger gV_{\mathcal G}(\vk_j)]}
{|\det[V_{\mathcal G}(\vk_i)^\dagger gV_{\mathcal G}(\vk_j)]|}.
\end{equation}
Under a basis rotation $V_{\mathcal G}(\vk_i)\to
V_{\mathcal G}(\vk_i)O_i$, with $O_i\in U(r)$, this link acquires the
phase $(\det O_i)^*\det O_j$, which cancels around each plaquette.
Using $U_{ij}^{\mathcal G}$ in Eqs.~\eqref{eq:plaquette-flux}
and~\eqref{eq:FHS} therefore gives the Chern number of the group,
independent of the basis chosen within the degenerate subspace.
\section{Magnon--phonon anticrossings}
\begin{figure}[htbp]
\centering
\includegraphics[width=.84\linewidth]{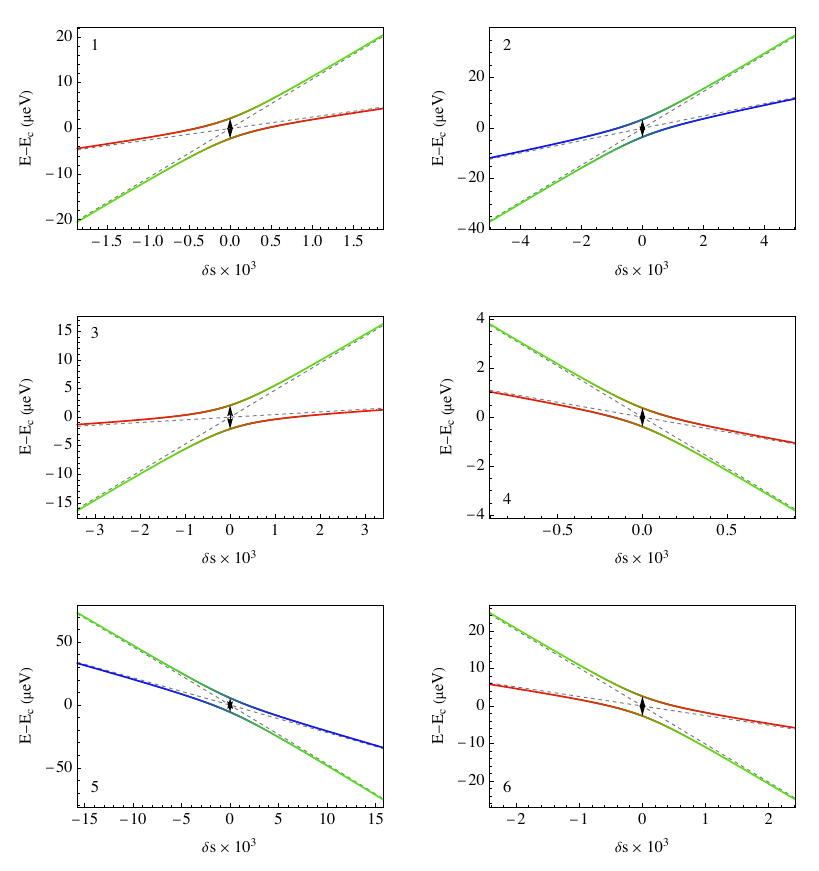}
\caption{Enlarged anticrossings 1--6 marked in Fig.~\ref{fig:main-spectra}. Solid colored curves show the coupled modes;
gray dashed curves show the uncoupled magnon and phonon branches.}
\label{fig:supp-gaps}
\end{figure}

Figure~\ref{fig:supp-gaps} resolves the six magnon--phonon
anticrossings marked in Fig.~\ref{fig:main-spectra}, using the same parameters.

\end{document}